\documentclass[apj,useAMS,usenatbib,usegraphicx,upmath]{emulateapj}
\usepackage{amsmath} \usepackage{amsfonts} \usepackage{amssymb}
\usepackage{apjfonts} \usepackage{color} \usepackage{graphicx}

\def\gsim{\;\rlap{\lower 2.5pt\hbox{$\sim$}}\raise 1.5pt\hbox{$>$}\;}
\def\lsim{\;\rlap{\lower 2.5pt\hbox{$\sim$}}\raise 1.5pt\hbox{$<$}\;}
\def\del{{\partial}}
\def\grad{\mbox{\boldmath{$\nabla$}}}

\def\bfzeta{\mbox{\boldmath{$\zeta$}}}

\def\tauth{\tau_{\rm th}}
\def\tin{\mathfrak{n}}

\shorttitle{RELAXATION TIME AND DISSIPATION INTERACTION}
\shortauthors{THRASTARSON \& CHO}
\submitted{}

\begin{document}

\title{Relaxation Time and Dissipation Interaction in Hot Planet
  Atmospheric Flow Simulations}

\author{Heidar Th. Thrastarson and James Y-K.\ Cho}

\affil{Astronomy Unit, School of Mathematical Sciences, Queen Mary
  University of London, Mile End Road, London E1 4NS, UK}

\email{H.Thrastarson@qmul.ac.uk; J.Cho@qmul.ac.uk}

\begin{abstract}

  We elucidate the interplay between Newtonian thermal relaxation 
  and numerical dissipation, of several different origins, in flow
  simulations of hot extrasolar planet atmospheres.  Currently, a
  large range of Newtonian relaxation, or ``cooling'', times ($\sim$10
  days to $\sim$1 hour) is used among different models and within a
  single model over the model domain.  In this study we
  demonstrate that a short relaxation time (much less than the
  planetary rotation time) leads to a large amount of unphysical,
  grid-scale oscillations that contaminate the flow field.  These
  oscillations force the use of an excessive amount of artificial
  viscosity to quench them and prevent the simulation from ``blowing
  up''.  Even if the blow-up is prevented, such simulations can be highly 
  inaccurate because they are either severely over-dissipated or 
  under-dissipated, and are best discarded in these cases.
  Other numerical stability and timestep size enhancers 
  (e.g., Robert-Asselin filter or semi-implicit time-marching schemes) 
  also produce similar, but less excessive, damping.  
  We present diagnostics procedures to choose the ``optimal'' simulation 
  and discuss implications of our findings for modeling hot extrasolar
  planet atmospheres.
                                                                      
\end{abstract}


\section{Introduction}\label{sec:intro}

There are many studies using a ``general circulation model'' (GCM) to
investigate the flow and temperature structure of close-in extrasolar
planet atmospheres
\citep[e.g.,][]{Showman02,Choetal03,Cooper05,Langton07,Choetal08,
  Dobbs-Dixon08,Showmanetal08a,Menou09,ThraCho10a,Rauscher10}.  GCMs
are advanced numerical models that solve a set of coupled, nonlinear
partial differential equations for the large-scale motions of a
shallow fluid on a rotating sphere.  In these sophisticated models,
numerous parameters are needed to specify the representation of
heating and cooling in the atmosphere and to stabilize the numerical
integration.

Thus far, not much emphasis has been given to the numerical aspects of
simulations in the extrasolar planet literature, in particular their
influence on the accuracy of the model results.  In an earlier paper,
\citet{ThraCho10a} has investigated the sensitivity of initial
condition on the extrasolar planet atmosphere flows.  In this work,
the focus is on another significant aspect---the subtle, and not so
subtle, interplay between numerical and physical parameters.  It
should be noted that, while the discussion is basically numerical,
this work is relevant to both theoretical studies and observations of
extrasolar planets.

GCMs usually solve the hydrostatic primitive equations \citep[see,
e.g.,][]{Salby96}, which filter sound waves so that only two important
classes of waves remain---Rossby, or planetary, waves (which evolve on
slow time scales) and gravity waves (which generally evolve on time
scales much shorter than the Rossby waves).  The spatial scales of the
two classes of motions are generally large and small, respectively.
Nonlinear advection, which has often been used to define a time scale
in extrasolar planet work so far, has roughly the same time scale as
the Rossby waves.  Generally, the amplitude of gravity waves, when
averaged over the globe, is very small compared to that of Rossby
waves, and most of the kinetic energy is contained in the large-scale,
slow motions.

A long-standing challenge in GCM theory is finding ways to deal with
fast waves accurately and efficiently.  The fast motions not only
force small timesteps to be taken (increasing the ``wall time'' of the
simulations), they also degrade the fidelity with which the equations
are solved.  Moreover, the very inaccuracy often causes the calculation
to ``blow up'' (become unstable), preventing any solution at all.
With certain types of numerical algorithms, such as implicit or
semi-implicit time-integration schemes, the timestep size restriction
can be alleviated.  But, artificial viscosity and various filters are
still required to stabilize the integration in general.

It is well known that, in conjunction with coarse resolution,
dissipation and filters can produce results that are seductively
misleading---even to the wary modelers.  For example, in the classic
Held-Suarez test for the dynamical core of GCMs for the Earth
\citep{HeldSuar94}, increasing the resolution generally leads to
enhanced equatorward shift of wave activity \citep{Wanetal08}.  The
shift becomes more evident in the simulations with horizontal
resolutions $\gsim\!$ T85 resolution (i.e., 85 sectoral and 85 total
modes) so that precise jet positions, for example, cannot be
ascertained at lower resolutions.  This is a relatively mild example,
but it is telling: the more extreme forcing condition for extrasolar
planets, it can be argued, will lead to larger or more sensitive
variations, given that the models have been designed and tested
for conditions appropriate for solar system planets.  In this
backdrop, even inter-comparing different GCMs for extrasolar planet
work becomes non-trivial.

In this paper we present and discuss examples of interesting behavior
when a GCM is stressed to its limits, with what may be considered a
typical hot, spin-orbit synchronized extrasolar planet condition.  
The implications are broad in the sense that the lessons are not just
limited to studies using GCMs, but also other types of global
circulation models.  The issues are present in all of them.

The basic plan of the paper is as follows.  In
Section~\ref{sec:method} we describe the GCM model we use and its
setup for the simulations described in this work.  In
Section~\ref{sec:spatial} we focus on the interaction between
artificial viscosity and the thermal relaxation time, which is an
important parameter in the representation of thermal forcing commonly
used in current studies.  In Section~\ref{sec:temporal} we examine
sensitivity of the simulations to the Robert-Asselin filter, which is
used to stabilize the time-marching scheme.  We conclude in
Section~\ref{sec:concl}, summarizing this work and discussing its
implications for extrasolar planet circulation modeling.

\section{Method}\label{sec:method}

\subsection{Governing Equations}\label{subsec:eqs}

In this work, we solve the same equations as in \citet{ThraCho10a}.
Here we briefly summarize the relevant aspects for the reader.  The
horizontal momentum equations are solved in the vorticity-divergence 
form:
\begin{eqnarray}\label{eq:vortdiv}
  \frac{\del \zeta}{\del t}\ & = &\
  {\bf k}\cdot\grad \times {\bf n} + {\cal D}_{\zeta}\\
  \frac{\del \delta}{\del t}\ & = &\
  \grad \cdot {\bf n} + \grad^2(E+\Phi) + {\cal D}_{\delta},
\end{eqnarray}
where $\bfzeta = \zeta\,{\bf k} = \grad\times{\bf v}$ is the
vorticity, $\delta = \grad\cdot{\bf v}$ is the divergence, ${\bf v}$
is the horizontal velocity, ${\bf k}$ is the vertical unit vector, $E
= ({\bf v}\cdot{\bf v})/2$, $\Phi$ is the geopotential, and
\begin{eqnarray*}
  {\bf n}\ =\ -(\zeta +f)\,{\bf k}\times{\bf v}\, -\,
  \dot{\eta}\frac{\del{\bf v}}{\del \eta}\, -\, \frac{RT}{p}\grad p,
\end{eqnarray*}
where $f$ is the Coriolis parameter, $p$ is pressure, $T$ is 
temperature and $R$ is the specific gas constant. 
The vertical coordinate is a generalized pressure coordinate, 
$\eta = \eta(p,p_s)$, with $p_s$ the bottom surface pressure, 
and $\dot{\eta}\equiv{\rm D}\eta/{\rm D}t$ with
\begin{eqnarray*} 
  \frac{{\rm D}}{{\rm Dt}}\ \equiv\ \frac{\del}{\del t}\, +\,
  \mathbf{v}\!\cdot\!\grad\, +\, \dot\eta\frac{\del}{\del\eta}
\end{eqnarray*} 
the material derivative. Hydrostatic equilibrium is assumed:
\begin{equation}
  \frac{\del\Phi}{\del\eta}\ =\ -\frac{RT}{p}\frac{\del p}{\del\eta},
\end{equation}
and the ideal gas law, $p = \rho RT$, where $\rho$ is density, 
is taken as the equation of state. The mass continuity equation 
is integrated from the bottom ($\eta$ = 1) to the top surface, 
$\eta_{\rm t}$, using the boundary conditions $\dot{\eta}$ = 0 
at both the top and the bottom, which yields an evolution equation 
for $p_s$:
\begin{equation}
  \frac{\del p_s}{\del t}\ =\ \int_{1}^{\eta_{\rm t}} \grad\! \cdot\! 
  \left(\frac{\del p}{\del \eta} {\bf v}\right)\, d\eta
\end{equation}
Integration of the continuity equation from $\eta_{\rm t}$ to $\eta$ 
yields a diagnostic equation for $\dot{\eta}$:
\begin{equation}
  \dot{\eta}\frac{\del p}{\del \eta}\ =\ -\frac{\del p}{\del t}\, -\, 
  \int_{\eta_{\rm t}}^{\eta} \grad\! \cdot\! 
  \left(\frac{\del p}{\del \eta} {\bf v}\right)\, d\eta.
\end{equation}
The diagnostic equation for  $\omega\equiv{\rm D}p/{\rm D} t$ 
is then:
\begin{equation}
  \omega\ =\ {\bf v} \cdot \grad p\, -\,   
  \int_{\eta_{\rm t}}^{\eta} \grad\! \cdot\! 
  \left(\frac{\del p}{\del \eta} {\bf v}\right)\, d\eta.
\end{equation}
Finally, the energy equation is
\begin{equation}\label{eq:temp}
  \frac{{\rm D}T}{{\rm D} t}\, -\,\frac{\omega}{\rho\, c_p}\ = \
  \frac{\dot{q}_{\rm{\tiny net}}}{c_p}\, +\, {\cal D}_T,
\end{equation}
where $c_p$ is the specific heat at constant pressure.  
In the final formulation of the equations,
terms involving ${\bf v}$ are represented in terms of the transformed
velocity ${\bf v}cos\phi$, where $\phi$ is latitude, in order to avoid
discontinuities at the poles.  Also, the equations are formulated
using ln($p_s$) instead of $p_s$ to avoid aliasing problems.  
The ${\cal D}$ terms in the vorticity, divergence and energy
equations represent horizontal diffusion, discussed in the next
subsection.

\subsection{Numerical Algorithm}\label{subsec:numerical}

To solve the equations described in the preceding subsection, we use
the Community Atmosphere Model (CAM 3.0), described in
\citet{Collinsetal04} and \citet{ThraCho10a}.  CAM is a well-tested,
highly-accurate hydrodynamics model employing the pseudospectral
algorithm \citep{Orszag70,Eliasenetal70}.  

For problems not involving sharp discontinuities (e.g., shocks
or, in atmospheric dynamics problems, fronts) and irregular
geometry, the pseudospectral method is superior to the standard grid
and particle methods \citep[e.g.,][]{Canutoetal88}.  To equal the
accuracy of the pseudospectral method for a problem solved with the
computational domain decomposed into $N$ grid points, one would need a
$N^{\rm th}$-order finite difference or finite element method with an
error of $O(N\Delta x)$, where $\Delta x$ is the grid spacing and
$O(\cdot)$ is the asymptotic order \citep[e.g.,][]{Nayfeh73}.  This is
because as $N$ increases, the pseudospectral method benefits in two
ways.  First, $\Delta x$ becomes smaller, which would cause the error
to rapidly decrease even if the order of the method were fixed.
However, unlike finite difference and finite element methods, the
order is {\em not} fixed: when $N$ is doubled to $2N$, the error
becomes $O[(\Delta x)^{2N}]$ in terms of the new, smaller $\Delta x$.
Since $\Delta x$ is $O(1/N)$, the error for the pseudospectral method
is $O[(1/N)^N]$.

Significantly, the error decreases faster than any finite power of $N$
since the power in the error formula is always increasing as well,
giving an ``infinite order'' or ``exponential'' convergence.  This
advantage is particularly important when many decimal places of
accuracy or high resolution is needed.  Note that in the vertical
direction CAM uses a finite differencing scheme, as in most GCMs.

For the spherical geometry, the horizontal representation of an
arbitrary scalar quantity $\xi$ consists of a truncated series of
spherical harmonics,
\begin{eqnarray*}
  \xi(\lambda,\mu)\ =\ 
  \sum^M_{N(m)} \sum^{N(m)}_{n = |m|}\xi^m_n P^m_n(\mu)\,e^{im \lambda},
\end{eqnarray*}
where $M$ is the highest Fourier (sectoral) wavenumber included in the
east-west representation; $N(m)$, which can be a function of the
Fourier wavenumber $m$, is the highest degree of the associated
Legendre functions $P^m_n$; $\lambda$ is the longitude; and, $\mu
\equiv \sin\phi$.  The spherical harmonic functions,
\begin{eqnarray}\label{eqn:expansion}
  Y^m_n(\lambda,\mu)\ =\ P^m_n(\mu)\, e^{im\lambda},
\end{eqnarray}
used in the spectral expansion are the eigenfunctions
of the Laplacian operator in spherical coordinates:
\begin{eqnarray}
  \grad^2\,Y^m_n\ =\ -\left[\frac{n(n + 1)}{R_p^2}\right]\, Y^m_n,
\end{eqnarray}
where
\begin{eqnarray*}
  \grad^2\ =\ \frac{1}{R_p^2}\left\{\frac{\del}{\del\mu}
    \left[\left(1 - \mu^2 \right)\frac{\del}{\del\mu}\right]\, +\, 
    \frac{1}{1 - \mu^2}\frac{\del^2}{\del\lambda^2}\right\}
\end{eqnarray*}
and $R_p$ is the planetary radius.  The set, $\{Y^m_n\}$, constitutes
a complete and orthogonal expansion basis \citep{ByronFuller92}.

In the Navier-Stokes equations, the diffusion terms appear as the
Laplacian of the dynamical variables \citep{Batchelor67}.  In our
case, the diffusion is generalized to the following
``hyperdissipation'' form \citep[e.g.,][]{ChoPol96}:
\begin{equation}
  {\cal D}_\chi\ =\ \nu_{2\mathfrak{p}}\left[(-1)^{\mathfrak{p}+1}\grad^{2\mathfrak{p}} + 
    {\cal C}\right]\,\chi,
\end{equation}
where $\chi = \{\zeta,\delta,T\}$ and ${\cal C} =
(2/R_p^2)^{\mathfrak{p}}$ is a correction term added to the vorticity
and divergence equations to prevent damping of uniform rotations for
angular momentum conservation.  In the above form, the $\mathfrak{p} =
2$ case is sometimes referred to as {\it superdissipation}.
Hyperdiffusion is added in each layer to prevent accumulation of power
on the small, poorly-resolved scales and to stabilize the integration.

\citet{ChoPol96} describes the effects of various hyperviscosities
(i.e., different values of $\mathfrak{p}$).  As discussed in that
work, a rational procedure for estimating roughly the value of
$\nu_{2\mathfrak{p}}$ can be obtained in the following way.  To damp
oscillations at the smallest resolved scale (set by the truncation
wave number, $n_t$), by an $e$-folding factor in time $\tau_d$, one
requires that
\begin{equation}\label{eqn:dissip}
  \nu_{2\mathfrak{p}}\ =\ 
  O\left\{\frac{1}{\tau_d}
    \left[\frac{R_p^2}{n_t(n_t+1)}\right]^{\mathfrak{p}}\right\}.
\end{equation}
Thereafter, the optimal value of $\nu_{2\mathfrak{p}}$ is obtained by
computing the kinetic energy spectrum (see Section~\ref{sec:results}).
Note that the precise value is problem specific, and the procedure
just described should be performed for {\it each} problem---as has
been done in this work.

In numerical solutions of time-dependent equations, there are two main
ways of marching in time.  Explicit methods give the solution at the
next time level in terms of an explicit expression which can be
evaluated by using the solution at the previous timestep.  Implicit
methods, on the other hand, require solving a boundary value problem
at each timestep.  Explicit time differencing is a more
straightforward numerical approximation to the equations.  In our
model, the time-marching is effected using a {\it semi}-implicit
scheme, a mixture of the two methods commonly used in GCMs.  In this
scheme, the equations are split into nonlinear and linear terms,
symbolically written:
\begin{eqnarray}
    \frac{\del\Psi}{\del t}\ =\ \mathcal{N}(\Psi)\, +\, 
    \mathcal{L}(\Psi),
\end{eqnarray}
where $\mathcal{N}(\Psi)$ and $\mathcal{L}(\Psi)$ denote the nonlinear
and linear terms, respectively, and $\Psi$ is the state of a variable
in $\chi = \{\zeta, \delta, T\}$.  

For the nonlinear terms, an explicit leapfrog scheme is used.  
This is a second-order, three-time-level scheme.  Because a
second-order method is applied to solve a differential equation which
is first-order in time, an unphysical computational mode is admitted,
in addition to the physical one.  In simulations containing nonlinear
waves, the computational mode can amplify over time, generating a time
splitting instability \citep{Durran99}.  Robert (1966) and Asselin
(1972) designed a filter to suppress the computational mode---hence
the time splitting instability.  This filter is applied in the GCM
used in the present work.  It is applied at each timestep so that
\begin{equation}
  \bar{\Psi}^\tin\ =\ \Psi^\tin\, +\, 
  \epsilon\left(\bar{\Psi}^{\tin-1}-2\Psi^\tin+\Psi^{\tin+1}\right),
\end{equation} 
where $\Psi^\tin = \Psi(\tin\Delta t)$, an overbar refers to the
filtered state, and $\epsilon$ specifies the strength of the filter.
The filter results in strong damping of the amplitude of the spurious
computational mode.  However, it also introduces a second-order error
in the amplitude of the physical mode with high values of $\epsilon$,
as we discuss further in Section~\ref{sec:temporal}.

Some parts of the equations can be solved implicitly with advantage.
In particular, the linear parts that produce fast gravity waves are
treated implicitly in many GCMs, including the one used in this work.
This treatment allows a larger timestep to be used, as mentioned in
Section~\ref{sec:intro}.  However, it is also at the cost of degraded
accuracy \citep[e.g.,][]{Durran99}.

As can be seen, time-integration of the primitive equations is not a
straightforward matter, even with a relatively simple method like 
the leapfrog scheme.  The theoretical analysis of the scheme is equally
complex.  The stability of the combined, semi-implicit leapfrog scheme
has been examined by \citet{Simmons78}, particularly with respect to
the basic state temperature profile.  They find the isothermal basic
state distribution to be more stable than a spatially-varying
distribution, with the stability generally increasing with higher
basic temperature.  In the present work an isothermal basic state of
1400~K is used.

\subsection{Calculation Setup}\label{sec:setup}

In addition to tuneable parameters associated with the numerical
scheme, such as the ones mentioned in the preceding subsection, the
representations of physical processes also require specification of
parameters.  Many of these are as yet poorly constrained by
observations or unobtainable from first principles (see, e.g.,
discussions in \citet{Choetal08}, \citet{Showmanetal08b}, and
\citet{Cho08}).  One example is thermal forcing (i.e., heating and
cooling) due to the irradiation from the host star and radiative
processes in the planetary atmosphere, which is represented in an
idealized way currently in all extrasolar planet atmosphere
simulations.  Many crudely represent the forcing by Newtonian
relaxation, as in this work
\citep[e.g.,][]{Cooper05,Langton07,Showmanetal08a,Menou09,Rauscher10,
ThraCho10a}.

In this representation, the net heating term in
equation~(\ref{eq:temp}) is represented by
\begin{eqnarray}
  \frac{\dot{q}_{\rm{\tiny net}}}{c_p}\  =\ 
  -\frac{1}{\tauth}\left(T - T_e\right),
\end{eqnarray} 
where $T_e = T_e(\lambda,\phi,\eta, t)$ is the ``equilibrium''
temperature distribution and $\tauth$ is the thermal relaxation (drag
or ``cooling'') time.  The appropriate values to use for this relaxation 
time (as well as the equilibrium temperature distribution) are poorly 
known and a large range of values has been used in the extrasolar 
planet literature.  In several studies, very short relaxation times---even 
less than an hour---and large $T_e$ gradients have been used
\citep[e.g.,][]{Showmanetal08a,Rauscher10,ThraCho10a}.  This
represents a rather ``violent'' forcing on the flow, depending on the
initial condition.

In this work, both $\tauth$ and $T_e$ are prescribed and barotropic
(i.e., $\del/\del\eta = 0$) and steady (i.e., $\del/\del t = 0$).  As
in \citet{ThraCho10a},
\begin{equation}\label{eq:Te}
  T_e\ =\ T_m + \Delta T_e \cos\phi \cos\lambda\, ,
\end{equation}
with $T_m = (T_D + T_N)/2$ and $\Delta T_e = (T_D - T_N)/2$, 
where $T_D$ and $T_N$ are the maximum and minimum temperatures at the
day and night sides, respectively.  All the simulations described in this
paper have $T_D = 1900$~K, $T_N = 900$~K.  Other physical parameters
chosen are based on the close-in extrasolar planet, HD209458b (see
Table~\ref{tab:params}).

\begin{deluxetable}{llll}
  \tablecaption{{Physical Parameters}
    \label{tab:params}} 
  \tablehead{}
  \startdata
  Planetary rotation rate & $\Omega$ &  2.1$\times$10$^{-5}$  & s$^{-1}$ \\
  Planetary radius & $R_p$ &  10$^8$  & m \\ 
  Gravity & $g$  &  10  & m s$^{-2}$ \\ 
  Specific heat at constant pressure & $c_p$ & 1.23$\times$10$^4$ &
  J kg$^{-1}$ K$^{-1}$ \\
  Specific gas constant & $R$ &  3.5$\times$10$^3$ & J kg$^{-1}$ K$^{-1}$ \\
  \\
  Mean equilibrium temperature & $T_m$ &  1400 & K \\
  Equilibrium substellar temperature & $T_D$ &  1900 & K \\
  Equilibrium antistellar temperature & $T_N$ &  900 & K \\
  Initial temperature & $T_0$ & 1400 & K \\
 \enddata
\end{deluxetable}

The spectral resolutions in the horizontal direction for the runs
described in the paper are T85 and T21.  The number refers to the
maximum total wavenumber, $n_t = \max\{N(m)\}$, at which
expansion~(\ref{eqn:expansion}) is truncated (e.g., T85 $\
\Rightarrow\ n_t = 85$); ``T'' means the truncation is such that $M =
N$ in equation~(\ref{eqn:expansion}), a ``triangular truncation'' in
wavenumber space.  A T85 spectral resolution corresponds roughly to
800$\times$400 grid points in physical space of grid-based methods
(roughly 200$\times$100 for T21 resolution).  The vertical direction
is resolved by 26 coupled layers, with the top level of the model
located at 3~mbar.  The pressure at the bottom $\eta$ boundary is
initially 1~bar, but the value of the pressure changes in time.  The
entire domain is initialized with an isothermal temperature
distribution, $T_0 = T_m = 1400$~K.  The flow field is initialized with a
small, random perturbation; specifically, values of the ${\bf v}$
components are drawn from a Gaussian random distribution centered on
zero with a standard deviation of 0.05~m~s$^{-1}$.  The sensitivity to
initial flow is described in detail in \citet{ThraCho10a}.

\section{Results}\label{sec:results}

\subsection{Spatial Dissipation}
\label{sec:spatial}

\begin{figure*} 
  \centerline{\includegraphics[width=18cm,height=23cm]{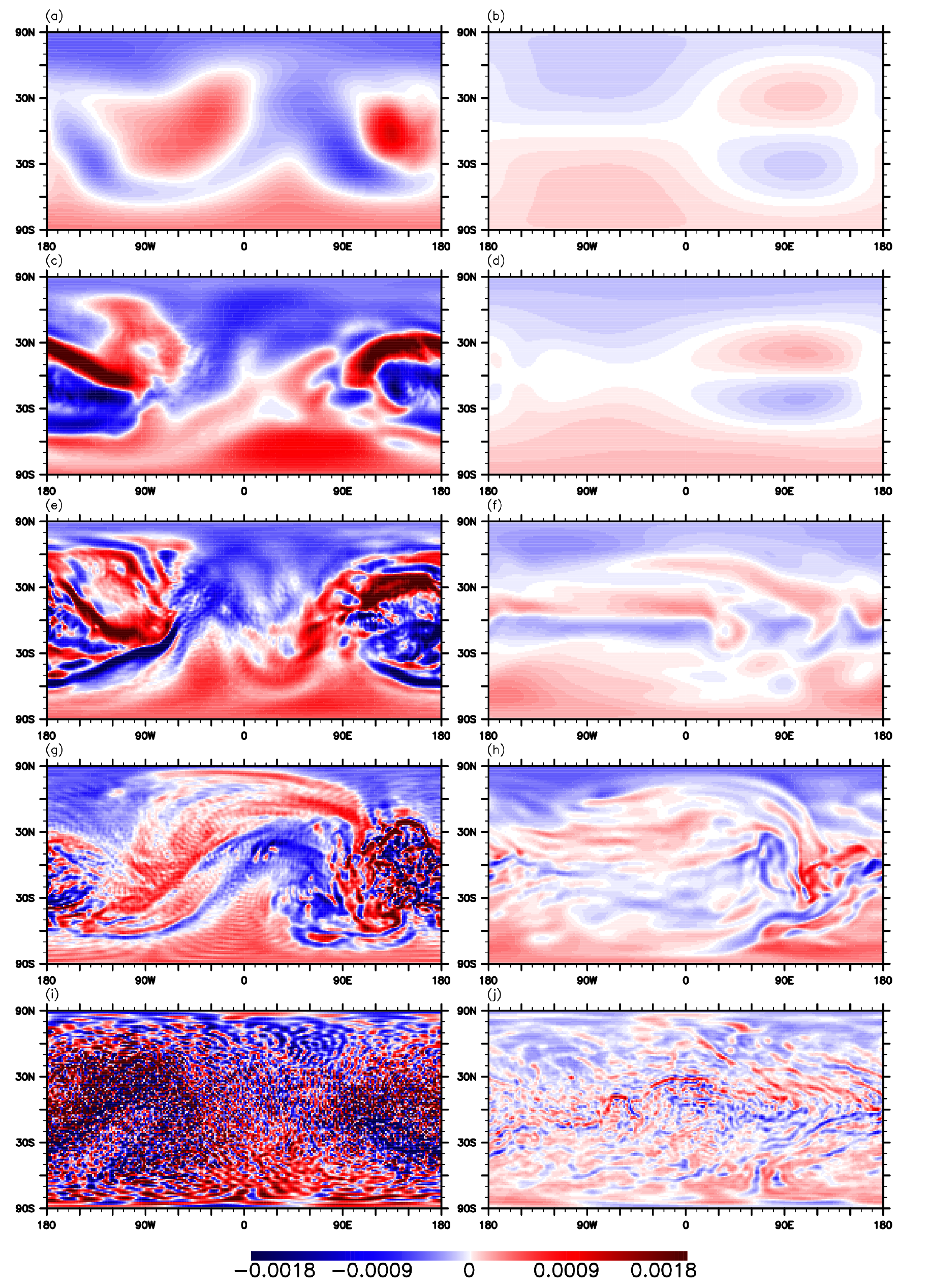}}
  \caption{Vorticity field at $t = 80\,\tau_p$ (planet rotations), near the 
  $p \approx 85\!$~mb level, for two sets of five simulations (left 
  column and right column) that are set up identically, except for the 
  viscosity coefficient and the thermal relaxation time.  The global 
  kinetic energy time series for all the runs have reached
  stationary (``equilibrated'') states and do not  qualitatively change for 
  $\sim\! 300\,\tau_p$.  The superdiffusion coefficient is 
  $\nu_4 = \{10^{24},10^{23}, 10^{22},10^{21},10^{20}\}$~m$^4$~s$^{-1}$, 
  decreasing from top to bottom in each column.  The panels in the left 
  column have a relaxation time of $\tauth = 0.1\,\tau_p$ while the panels 
  on the right have $\tauth = 3\,\tau_p$.  Red (blue) color represents 
  positive (negative) values of vorticity, with units $s^{-1}$.
  \label{fig:diss-tau}}
\end{figure*}

Table~\ref{tab:sims-nu} lists all the runs discussed in this
subsection.  Simulations are performed with the setup described above,
but with varying strength of artificial viscosity ($\nu$ and
$\mathfrak{p}$) and the forcing timescale ($\tauth$).
Figure~\ref{fig:diss-tau} presents the relative vorticity field near
the $p \approx 85\!$~mb level; this is approximately a quarter of the
way down from the top of the computational domain.  The field at $t =
80\,\tau_p$ is shown in cylindrical equidistant projection, centered
at the equator, for ten simulations in which the setup is identical
except for the values of $\tauth$ and $\nu$ ($\mathfrak{p} = 2$,
superdissipation); here, $\tau_p = 2\pi/\Omega$ is the planetary
rotation period.
Positive vorticity (red color) signifies local rotation in the same
direction as the planetary rotation (counter-clockwise in the northern
hemisphere), and opposite for the negative vorticity.  The panels on
the left column all have the same short value of $\tauth =
0.1\,\tau_p$, while the panels on the right column all have $\tauth =
3\,\tau_p$ for five different values of $\nu$.  In all the runs shown,
the global kinetic energy time series have reached stationary
(``equilibrated'') state and do not change qualitatively for
approximately 300$\,\tau_p$.

For a given value of $\tauth$, simulations with different $\nu$'s
generally share some common features over a range of $\nu$'s.  But,
there are clear differences in the character of the flow and
temperature fields.  The differences, which are both qualitative and
quantitative, arise from the strength of dissipation.  Moreover, $\nu$
can affect the temporal behavior as well.  For example, temporal
variability can be muted with larger $\nu$.  Not surprisingly, in the
strongest dissipation cases [panels~(a) and (b)] variability in time
is essentially completely quenched and the flow structures are quite
smooth in appearance.  These are examples of runs which are severely
over-dissipated.

At the other extreme, runs can also be severely under-dissipated.
This is shown in panels~(i) and (j) in Figure~\ref{fig:diss-tau}.
Note that the common $\nu$ value in these runs is four orders of
magnitude smaller than that for the runs of panels~(a) 
and (b).
A quick visual check of panels~(i) and (j) immediately shows the
physical fields dominated by small-scale oscillations: this is
numerical noise.  Here, by ``small'' we mean scales near the
grid-scale, $l = O(\Delta x)$.  Typically, runs like these blow
up---or at least they should (see Section~\ref{sec:temporal}).
Simulations often blow up long before the small-scales contain any
significant amount of energy compared to the large-scales.  As we
discuss more later, this is because the calculation {\em correctly}
becomes unstable.  But, sometimes misbehaving simulations can be
surprisingly resilient and not crash.  This is usually a signal that
bad numerics is at play.

\begin{deluxetable}{cccc} 
  \tablecaption{{List of Runs Discussed}
    \label{tab:sims-nu}} 
  \tablehead{Run & $\nu_{2\mathfrak{p}}$\ [m$^4$~s$^{-1}$]  & 
    $\mathfrak{p}$ & $\tauth/\tau_p$}
   \startdata
  N1a & 1$\times$10$^{24}$   & 2 &  .1  \\
  N1b & 1$\times$10$^{24}$   & 2 &   3  \\
  N2a & 1$\times$10$^{23}$   & 2 &  .1  \\
  N2b & 1$\times$10$^{23}$   & 2 &   3  \\
  N3a & 1$\times$10$^{22}$   & 2 &  .1  \\
  N3b & 1$\times$10$^{22}$   & 2 &   3  \\
  N4a & 1$\times$10$^{21}$   & 2 &  .1  \\
  N4b & 1$\times$10$^{21}$   & 2 &   3  \\
  N5a & 1$\times$10$^{20}$   & 2 &  .1  \\
  N5b & 1$\times$10$^{20}$   & 2 &   3  \\
  N6a & $^\dagger$6$\times$10$^{12}$ &  1 &  .1 \\
  \enddata
  \tablecomments{$\nu$ is the hyperviscosity coefficient 
    and $\mathfrak{p}$ the order index of the hyperviscosity. 
    $\tauth$ is the thermal relaxation timescale and
    $\tau_p$ is one planet rotation.    
    All the simulations are 
    run at T85 resolution with a timestep $\Delta t$ of 60 s and a 
    Robert-Asselin filter coefficient $\epsilon$ of 0.06.\\
    $^\dagger$The units for this $\nu$ are [m$^2$~s$^{-1}$].}
\end{deluxetable}

As expected, increasing $\nu$ leads to decreasing small-scale
oscillations and to increasingly smoother fields.  However,
significantly, we note that for a given value of $\nu$ for the two
$\tauth$'s (cf., panels of the same row in Figure~\ref{fig:diss-tau})
shorter $\tauth$ in a run admits much more pronounced grid-scale
oscillations.  For example, with $\nu_4 = 10^{22}$~m$^4$~s$^{-1}$
[panels (e) and (f)], the viscosity is clearly {\it insufficient} to
suppress small-scale oscillations in the case of $\tauth =
0.1\,\tau_p$, while no small-scale oscillations are present in the
calculation with longer $\tauth$.  More importantly, a $\nu$ value
which appears to be acceptable for the shorter relaxation time [e.g.,
panel~(a)] is clearly over-dissipative for the run with the longer
$\tauth$ [e.g., panel~(b)]: here, the calculation in (b) should be
compared with that in (f), which is clearly a much less dissipated run
than that in (b).  Hence, running a simulation at a single
$\tauth$---even if $\nu$ were varied---would not produce trustworthy
results since the parameter space is at least two-dimensional.

The implication of this is serious.  In many current simulations of
hot planet atmospheric flows, a range of $\tauth$'s is specified,
spread over the model atmosphere domain, which always contains a
region with a short $\tauth$.  This forces those model atmosphere
calculations to be excessively noisy {\it and} excessively dissipated,
in different atmospheric regions of the computational domain.  Once
noise appears in the calculation somewhere in the domain, the entire
domain becomes quickly contaminated.  Note that an inherently smooth
field---such as temperature compared to vorticity, for example---would
not reveal the noise as well, since it is essentially two integrations
(a smoothing operation) of the vorticity field.  In other words,
temperature possesses a steep (narrow) spectrum like the stream
function, as opposed to a shallow (broad) spectrum like the vorticity.
Similarly, other averaging (integrating) procedures, such as taking
zonal (eastward) and/or temporal means, would obscure, possibly
mislead, the analysis of the simulation if unaveraged ``higher-order''
fields like vorticity are not considered concomitantly.

\begin{figure*} 
    \begin{center}
    \begin{tabular}{cc}
      \resizebox{9cm}{9cm}{\includegraphics{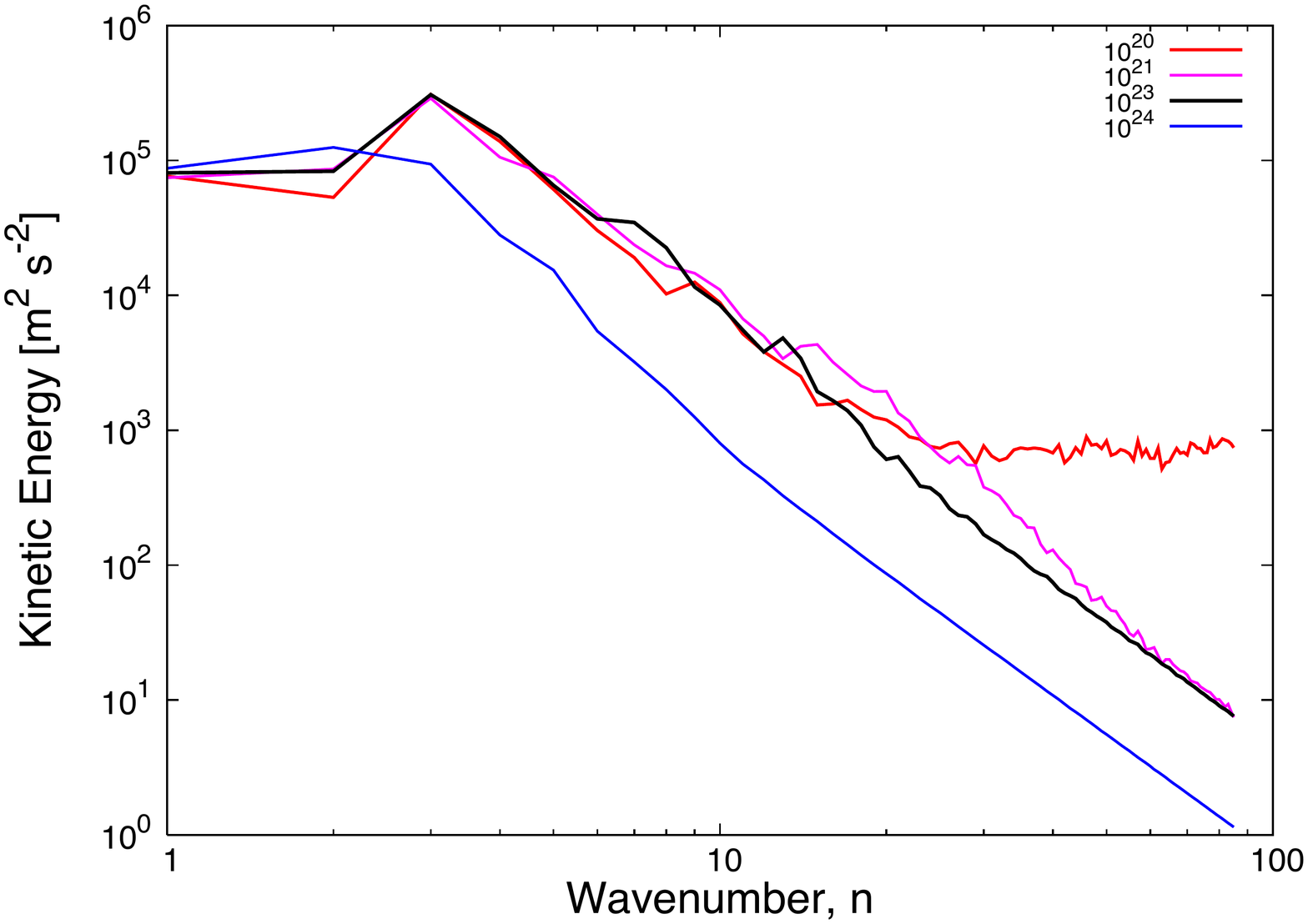}} &
      \resizebox{9cm}{9cm}{\includegraphics{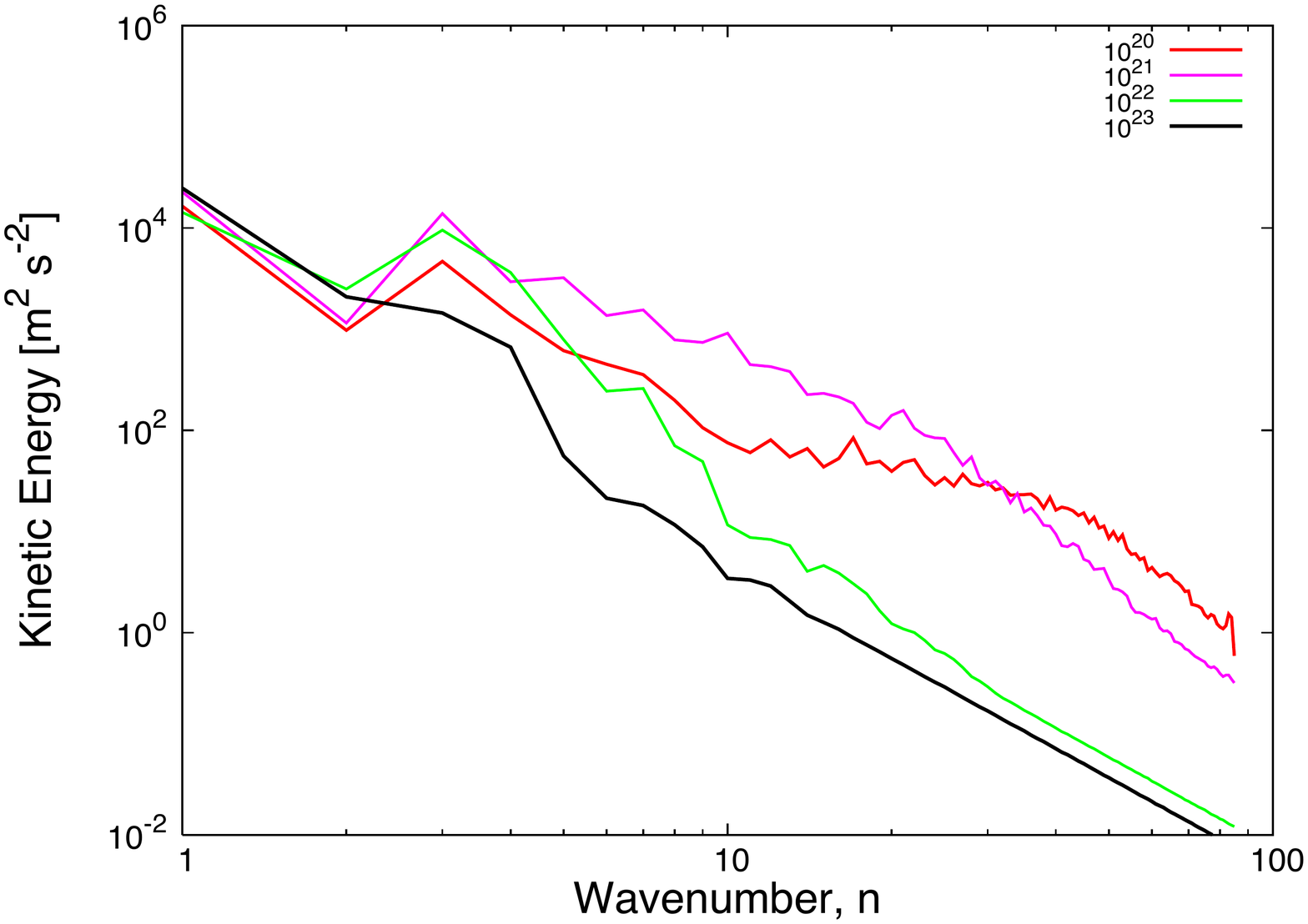}}
    \end{tabular} 
  \caption{Kinetic energy spectra for the fields shown in 
  Figure~\ref{fig:diss-tau} for simulations that are set up identically, 
  except for the artificial viscosity and the thermal relaxation time. 
  The runs shown in the left panel have a relaxation time of 
  $\tauth = 0.1\,\tau_p$ while the runs on the right panel have 
  $\tauth = 3\,\tau_p$.  The different colored lines are for different
   values of $\nu$, as indicated in the legend.
   Note the different scales on the two panels---much more 
   kinetic energy is contained in the flow when the relaxation time 
   is short.
   The spectra reveal both under-dissipated (e.g. red line, 
   left panel) and over-dissipated (e.g. blue line, left panel) flow fields. 
   A $\nu$-value that seems to give a reasonable spectrum for the 
   short $\tauth$ (e.g. black line, left panel) results in over-dissipation 
   for the longer $\tauth$ (black line, right panel).
  \label{fig:spec}}
  \end{center}
\end{figure*}

\begin{figure} 
  \centerline{\includegraphics[width=9cm,height=9cm]{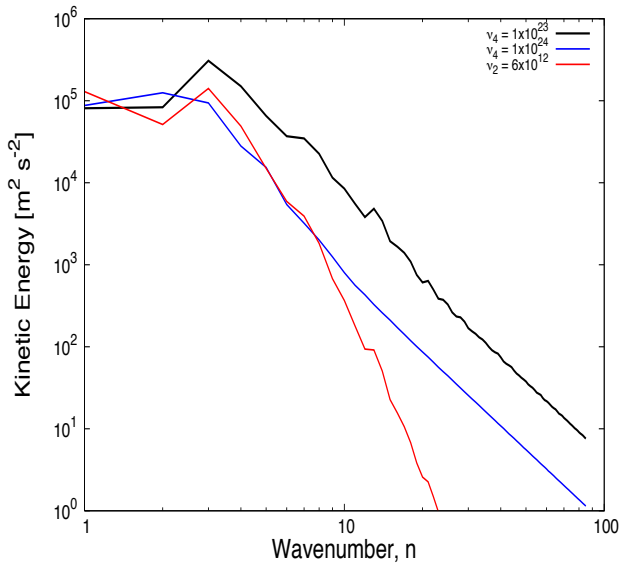}}
  \caption{Kinetic energy spectra for simulations that are set up
    identically, except for the artificial viscosity.  The different
    lines refer to different values of $\nu$, as indicated in the
    legend.  The blue and black lines are the same as in the left
    panel of Figure~\ref{fig:spec}, for which the viscosity is of
    biharmonic form ($\grad^4$ with $\mathfrak{p} = 2$).  But, the red
    line is for a simulation where the order of the viscosity is lower
    ($\mathfrak{p} = 1$), the normal Newtonian viscosity.
    \label{fig:del2}}
\end{figure}

We wish to emphasize here that, in contrast to what might be the
customary view, numerical noise and blow-ups are useful.  Simulations
with severe forcing should be {\it allowed} to crash---or at least
halted, when near grid-scale oscillations are visible in the flow
field.  Any phenomena observed thereafter would be seriously
compromised in accuracy, and quite possibly entirely artifactual
\citep{Boyd00}.  In numerical work, it is easy to get lured into
believing a calculation by not heeding important telltale signs.

There is a rational way to diagnose the onset of the small-scale error
sources---as well as the excessive dissipation---in a simulation.
This is illustrated in Figure~\ref{fig:spec}, which contains the
kinetic energy spectra of the fields presented in
Figure~\ref{fig:diss-tau}.  To the best of our knowledge, this is the
first time kinetic energy spectra are shown in extrasolar planet
atmosphere flow simulations.  They provide an important diagnostic,
when used in conjunction with instantaneous fields (see, e.g.,
\citet{ChoPol96} and \citet{Koshyk99} for a discussion of kinetic
energy spectra and horizontal diffusion), and can be used to choose an
appropriate $\nu$ value.

The left set of spectra in Figure~\ref{fig:spec} corresponds to runs
with the shorter $\tauth~=~0.1\,\tau_p$ in the left column of
Figure~\ref{fig:diss-tau}, and the right set of spectra in
Figure~\ref{fig:spec} corresponds to runs with the longer
$\tauth~=~3\,\tau_p$ in the right column of Figure~\ref{fig:diss-tau}.
Visual inspection of the vorticity fields along the left column of
Figure~\ref{fig:diss-tau} suggests the runs in panels~(a) and (c) are
not much affected by the small-scale oscillations [if at all in the
run of panel~(a)].  This can be quantified by confirming that the
corresponding spectra in Figure~\ref{fig:spec} (left panel) are the
blue and black lines (runs N1a and N3a, respectively).  In fact, the
blue line clearly reveals a case of over-damping, in which all scales
are less energetic than the corresponding scales in the other runs.

In contrast, note the appearance of near-grid-scale waves in physical
space, for the run in panel~(i) in Figure~\ref{fig:diss-tau},
indicated by a tendency for the spectrum (red line in left panel of
Figure~\ref{fig:spec}) to peel off and curl up near---and
considerably to the left (larger scale) of---the aliasing limit;
this is
\begin{eqnarray*}
  n_a\ =\ \frac{2\pi R_p}{3\Delta x},
\end{eqnarray*}
which is $\approx 85$ in our case, since $\Delta x$ is chosen to be
``alias-free'' up to $n_t$ \citep{Orszag71}.  Clearly, our de-aliasing
procedure, of inverse transforming onto a physical grid that is $3n_t
+ 1$ around the longitude, is not successful in runs N5a and N4a, as
well as in run N3a (spectrum not shown).  This is because increasingly
greater resolution is needed as the calculation proceeds, as discussed
below.  In turbulence simulations, this peeling off behavior is known
as an ``energy pile-up'' or ``spectral blocking'' (because direct
energy cascade to high wavenumbers in three-dimensional turbulence is
blocked).  It is not limited to spectral methods.  It is universal to
all methods which discretize space.

Spectral blocking can cause numerical instability in the time
integration of any nonlinear equations.  The instability arises due to
the quadratically nonlinear term in the solved equations.  For example, 
a typical quadratically nonlinear term (in one-dimensional Cartesian
geometry for simplicity) gives:
\begin{eqnarray*}
  \psi\frac{\del\psi}{\del x} &\  =\ &     
  \left( \sum^K_{p = -K} a_p\, e^{ipx}\right) 
  \!\cdot\!\left(\sum^K_{q = -K} i\, q\, a_q\, e^{iqx} \right) = \sum^{2K}_{k = -2K} b_k\, e^{ikx}.
\end{eqnarray*}
Here, $\psi(x,t)$ is an arbitrary one-dimensional scalar function,
which is Fourier expanded; $b_k$ are given by a sum over the products
of the $a_k$; $K$ is the truncation wavenumber, corresponding to $n_t$
in equation~(\ref{eqn:dissip}).  Note that the nonlinear interaction
generates high wavenumbers, $k > K$, which will be aliased into
wavenumbers on the range $k\in [-K, K]$.  This induces an
unphysical inverse cascade of energy from high wavenumbers to low
wavenumbers.

It is important to realize that the above cascade injects artificial
energy into {\it all} scales.  The injection is simply more noticeable
in the small scales since not much energy is contained there in the
absence of blocking.  Oscillations of size $l = O(\Delta x)$ are a
precursor to breakdown of computational fidelity.  These oscillations
are insidious because they require higher and higher resolution in the
calculation over time.  Without the increasing resolution, they
deteriorate the accuracy of the simulation on all scales as the
calculation proceeds, as pointed out in \citet{ThraCho10a}.  Although
some blocking is almost inevitable in a long time integration of a
nonlinear system (unless the dissipation is unrealistically large), it
can be monitored and controlled---albeit better in some methods than
in others.

The left and right panels of Figure~\ref{fig:spec} reveal not only how
the appropriate dissipation can be chosen, but also the crucial
interplay between the small-scale noise and $\tauth$---hence,
underscoring the importance of using {\it both} the spectra and the
physical field in analyzing a calculation.  Consider, for example, the
``optimal'' calculation (i.e., least affected by too much or too
little dissipation) for the short $\tauth$ runs.  The calculation with
$\nu_4 = 10^{23}$~m$^4$~s$^{-1}$ (black line in the left panel) is
devoid of non-physical build up of energy at the smallest scales while
still retaining the same amount of energy in the large scales as in
the calculations with smaller $\nu$.  On the other hand, the
calculation with longer $\tauth$ but same $\nu$ (black line in the
right panel) is clearly over-dissipated, containing less energy
compared to the other calculations on essentially all the scales.
Hence, if the $\nu$ value were ``tuned'' with the calculations with
shorter $\tauth$ (only), then a calculation with a different $\tauth$
(say a longer one, as in this example) would be over-damped.  In other
words, a correct $\nu$ value cannot be obtained independent of
$\tauth$.  Actually, $\nu = \nu\,(\tauth,\epsilon,\cdots)$, where
``$\cdots$'' includes $T_e$, $R_p$. semi-implicitness, etc.

The above behavior is generic.  Simulations performed with a greater
range of $\tauth$ (down to 0.01\,$\tau_p$) and $\nu$ and
$\mathfrak{p}$, exhibit the same basic behavior; and, it is present
throughout the model domain; grid-scale oscillations can appear in the
duration of a calculation anywhere in the domain.  These oscillations
can be controlled to some degree in mild cases, as outlined above.
However, grid-scale oscillations are dominant near the top of the
domain for {\it all} values of $\nu$ considered.  In this situation,
it is common in GCM studies to include a ``sponge layer", where
dissipation is artificially enhanced in the topmost layers.  While
this can damp unphysical oscillations, it can also have spurious
effects.  The effects of ``sponges" as well as other boundary
conditions will be described elsewhere.

Figure~\ref{fig:del2} shows how the spectrum is affected when the form
of the artificial viscosity is of lower order.  The blue and black
lines (runs N1a and N2a, respectively) are the same as in the left
panel of figure Figure~\ref{fig:spec}. They can be compared to the red
line (run N6a), which shows the spectrum from a simulation that is
identical to the other two runs in the figure, except for the value of
$\nu$ and the order of the viscosity operator (here $\mathfrak{p} =
1$).  In this case, the energy in the small scales (high wavenumbers)
is dissipated much more strongly.  More importantly, essentially {\it
  all} wavenumbers are affected by the lower order viscosity;
and, as discussed in \citet{ChoPol96}, the slope of the spectrum
becomes steeper---even at wavenumbers well below the truncation scale.

\subsection{Temporal Dissipation}
\label{sec:temporal}

If the solved equations support several types of waves, as with the
primitive equations, the maximum stable timestep is limited by the
Courant number,
\begin{eqnarray*} \mu^* \equiv c_{\rm
    max}\left(\frac{\Delta t}{\Delta x}\right),
\end{eqnarray*}
where $c_{\rm max}$ is the maximum horizontal wind speed associated
with the fastest propagating wave.  Some fast waves are of little
physical significance, but they enslave $\Delta t$ to be small.
Implicit schemes do permit a larger timestep size to be used than in
explicit schemes, often making the former more computationally
efficient.  However, for nonlinear equations, implicit schemes have a
high cost per timestep because a nonlinear boundary value problem must
be solved at each timestep.

As noted, a semi-implicit algorithm is commonly used in GCMs.  In
general, the implicit and explicit parts in the algorithm may be of
same or different order.  Treating some terms explicitly while others
implicitly may appear strange, but there are some major advantages.
First, because the nonlinear terms are treated explicitly, it is only
necessary to solve a {\it linear} boundary value problem at each timestep.  
Second, the hyperdissipation terms, which involve even number
of derivatives, impose a much stiffer timestep requirement than the
advective terms; for example, $\Delta t$ is $O(1/N^4)$ and $O(1/N^2)$,
respectively, for the Newtonian viscosity ($\mathfrak{p} = 1$).
Hence, the semi-implicit algorithm stabilizes the most unwieldy terms.
Third, in general circulation and other fluid dynamics problems,
advection is crucial; therefore, it is important to use a high order
time-marching scheme with a short timestep to accurately compute
phenonmena or structures such as frontogenesis, advection of storm
systems, and turbulent cascades.  There is little advantage in
treating the nonlinear terms implicitly because a timestep longer than
the explicit advective stability limit would be too inaccurate.

Note that, although it is possible to treat the time coordinate
spectrally, it is generally more efficient to apply spectral methods
to the spatial coordinates only because time marching is usually much
cheaper than computing the solution simultaneously over all
space-time.  In general, much less concern is given to the temporal
accuracy than the spatial accuracy of GCMs---usually with good
justification: spatial errors pose greater problems, especially for
the short and medium range duration runs typically performed with the
models.  This obviously does not apply for long duration runs,
particularly if quantitative predictions are sought \citep{ThraCho10a}.

\begin{deluxetable}{clcl} 
  \tablecaption{{Summary of Runs for Time Filter Sensitivity}
    \label{tab:sims-eps}} 
  \tablehead{ Run &  $\ \ \ \epsilon$  & 
    $\tauth/\tau_p$  & Notes } 
   \startdata
  E1a & 0.001  &  .1  &  blow-up ($t = 1\,\tau_p$) \\
  E1b & 0.001  &  3  &  blow-up ($t = 71\,\tau_p$)  \\
  E2a & 0.002  &  .1  &  blow-up ($t = 1\,\tau_p$)  \\
  E2b & 0.002  &  3  &  \\
  E3a & 0.006  &  .1  &  blow-up ($t = 8\,\tau_p$) \\
  E3b & 0.006  &  3  &  \\
  E4a & 0.01  &  .1  &  \\
  E4b & 0.01  &  3  &  \\
  E5a & 0.06  &  .1  &  \\
  E5b & 0.06  &  3  &  \\
  E6a & 0.1  &  .1  &  \\
  E6b & 0.1  &  3  &  \\
  \enddata
  \tablecomments{$\tauth/\tau_p$ is the thermal relaxation time in
    units of planetary rotations, and $\epsilon$ is the Robert-Asselin
    filter coefficient.  All the runs are at T21 resolution and have
    $\nu_4$ = 10$^{22}$~m$^4$~s$^{-1}$.  The timestep is $\Delta t$ =
    240~s.}
\end{deluxetable}

As already discussed, a computational mode arises in the
leapfrog scheme, which is an example of a two-step scheme:
\begin{eqnarray}
  \Psi^{\tin + 1}\ =\ \Psi^{\tin - 1}\, + \,
  \mathcal{F}(\Psi^\tin, \mathbf{x}, t^\tin; \epsilon),
\end{eqnarray} 
where $\mathbf{x}\in\mathbb{R}^2$; recall that $\epsilon$ is the
Robert-Asselin time filter coefficient.  Computational modes arise in
all multistep methods.  Fortunately, in some multistep methods $\Delta
t$ can be chosen to keep the amplitude of the modes from growing.
However, the leapfrog scheme is unstable for diffusion, for all
$\Delta t$.  For this reason, the diffusion part of the equations is
``time-lagged'' by evaluating the diffusion terms at the time level
$(\tin - 1)$.  This effectively time-marches the diffusion part by a
first-order scheme.

The Robert-Asselin filtered leapfrog scheme has been analyzed by
\citet{Durran91} for the simple oscillation equation,
\begin{eqnarray}
  \frac{{\rm d}\psi}{{\rm d}t}\ =\ i\omega\psi,
\end{eqnarray}
where $\omega$ is the frequency of oscillation.  That analysis shows
that, in the limit $\omega \Delta t \ll 1$, the relative phase-speed
error of the physical mode is
\begin{equation}\label{eq:phase}
  {\cal R}_{\rm phys}\ =\ 1 + 
  \left[\frac{1+2\epsilon}{6\,(1-\epsilon)}\right](\omega\Delta t)^2.
\end{equation}
Therefore, the phase of the numerical solution leads the actual
solution in time, and the error increases with larger $\epsilon$.  No
analysis exists to guide in choosing $\epsilon$.  Hence, it is
important to assess the sensitivity of the simulations to the filter.
In this work, we have performed series of simulations in which
$\epsilon$ has been varied while keeping everything else fixed, for
different values of relaxation time $\tauth$.  The simulation
parameters are summarized in Table~\ref{tab:sims-eps}.

\begin{figure} 
  \centerline{\includegraphics[width=9cm,height=9cm]{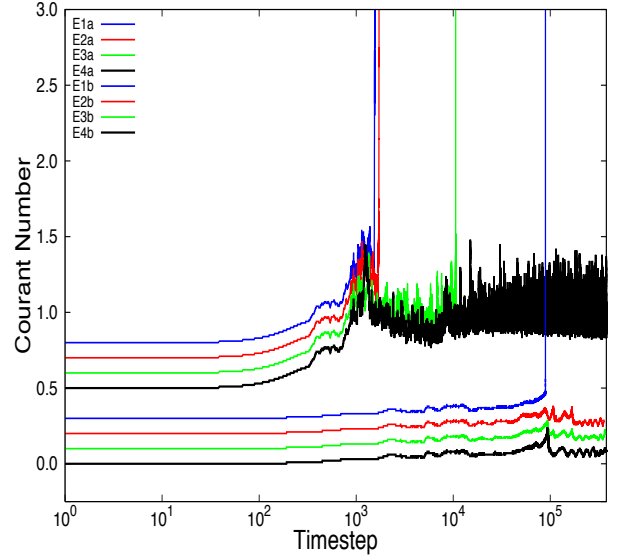}}
  \caption{Courant number as a function of time for two sets of
  runs with different values of $\tauth$ in each set.  The four runs 
  within each set have different $\epsilon$ values, setting the 
  strength of the Robert-Asselin time filter. For 
  clarity each time series in a set has been offset vertically by 0.1 in
  the plot; and, the two sets, as groups, have been offset vertically by
  0.5. The lower set of runs have $\tauth$ = 3 $\tau_p$, while the 
  upper set of runs have $\tauth$ = 0.1 $\tau_p$. For each set the run with 
  $\epsilon$ = 0.001 is indicated with a blue line, $\epsilon$ = 0.002 
  a red line, $\epsilon$ = 0.006 a green line, and $\epsilon$ = 0.01 a 
  black line.
\label{fig:cour}}
\end{figure}

Figure~\ref{fig:cour} shows the evolution of the Courant number
$\mu^*$ for simulations with different $\epsilon$, for two sets of
runs with different values of $\tauth$ in each set.  Note that for
clarity each time series in a set has been offset vertically by 0.1 in
the plot; and, the two sets, as groups, have been offset vertically by
0.5.  At the T21 resolution of the runs shown, for $\tauth =
3\,\tau_p$ and $\nu = 10^{22}$~m$^4$~s$^{-1}$, it is found that a
value of at least $\epsilon = 0.002$ is needed to prevent the
simulation from succumbing to time-splitting instability.  

With shorter relaxation time ($\tauth = 0.1\,\tau_p$), a {\it larger}
value of $\epsilon$ is required for the simulation to proceed without
blowing up; this is perhaps not surprising, in light of the preceding
discussion.  But, remarkably, even without explicit numerical
viscosity turned on (i.e., $\nu$ set to 0), the simulation can proceed
without crashing; and, this is so despite the fact that the physical
field is completely swamped with noise!  When $\tauth = 3\,\tau_p$,
runs do not crash as long as $\epsilon \ge 0.006$.  However, with
$\tauth = 0.1\,\tau_p$, the minimum $\epsilon$ for not crashing is an
order of magnitude greater.  Evidently, an $\epsilon$ value used in
earlier studies of Earth's atmosphere should be adjusted when adapting
an Earth GCM for extrasolar planet study.  In general, a shorter
$\tauth$ or lower viscosity requires stronger Robert-Asselin filtering
to prevent blow-up.

\begin{figure*} 
  \epsscale{1.} 
  \plotone{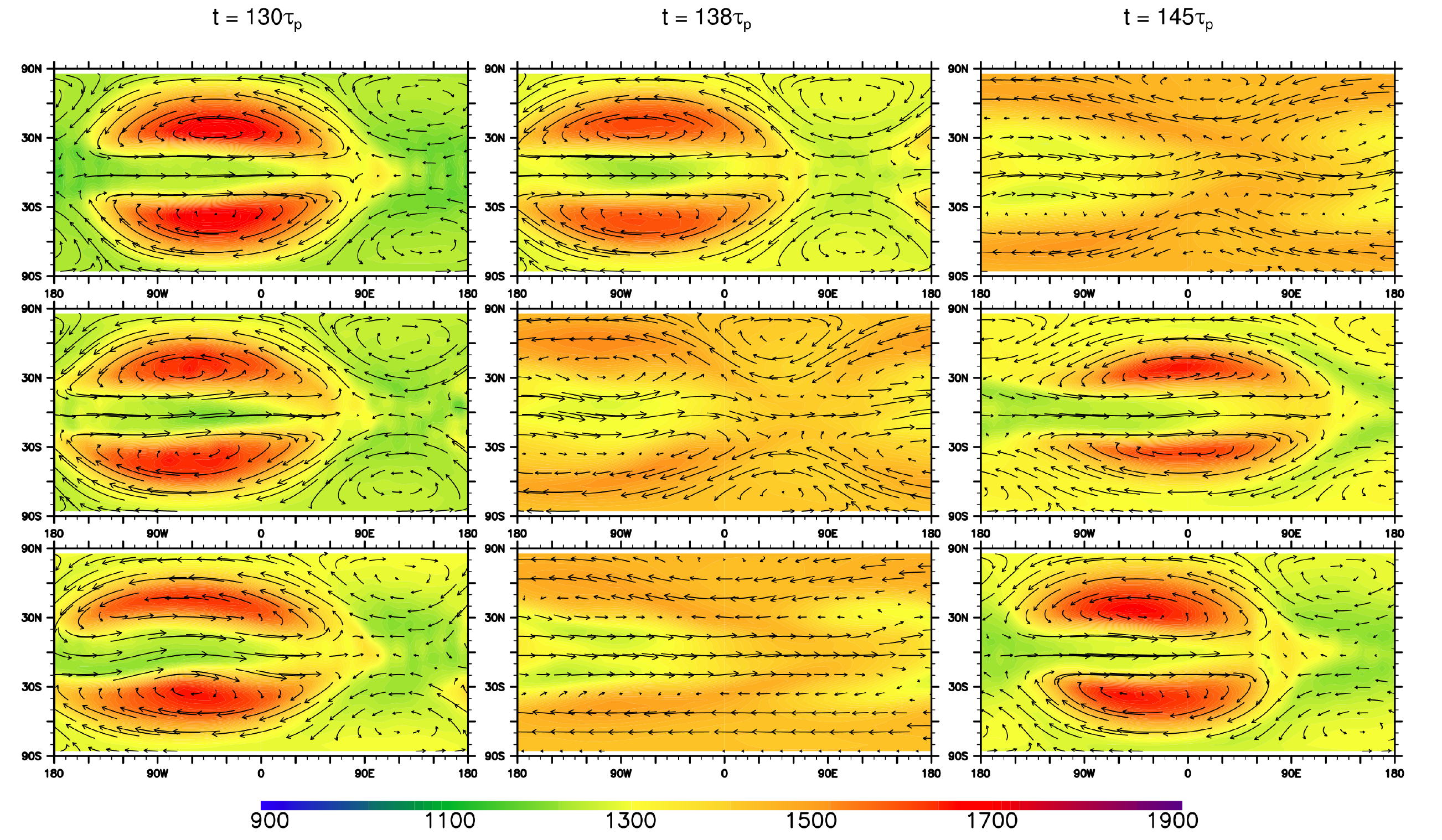} 
  \caption{Temperature (color coded in K) with streamlines overlaid, 
      for three simulations differing only in the value of 
      $\epsilon$, shown at three moments in time. From left to right, 
      the snapshots are taken at $t = \{130,138,145\} \tau_p$.  The 
      top row is from a run with $\epsilon$ = 0.01, the middle row 
      with $\epsilon$ = 0.06 and the bottom row $\epsilon$ = 0.10.
      All the fields are shown at the $p \approx 900\!$~mb level. 
      The substellar point is at $0^{\circ}$ longitude and latitude.
\label{fig:eps}}
\end{figure*}

\begin{figure} 
  \centerline{\includegraphics[width=9cm,height=9cm]{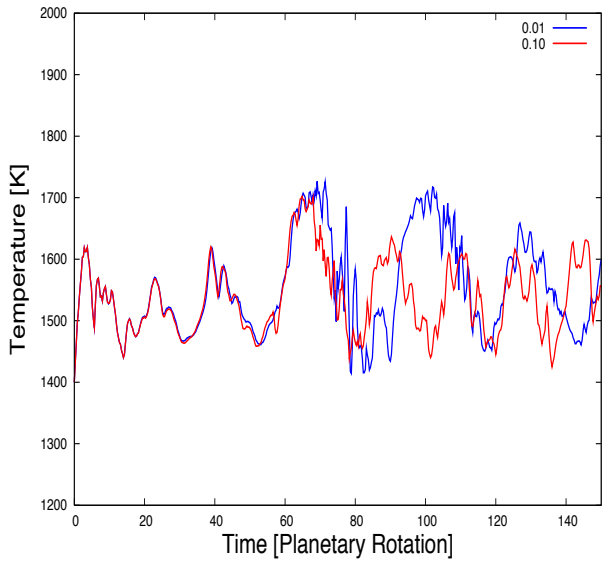}}
  \caption{Temperature at a fixed point, at 0$^{\circ}$ longitude and 
    30$^{\circ}$ latitude, as a function of time for the first 150 planet 
    rotations in two simulations. The two curves show results of 
    simulations that are identical apart only from the strength of the 
    Robert-Asselin time filter.  The red curve is from a run with 
    $\epsilon$ = 0.10, and the blue curve from a run with $\epsilon$ = 0.01. 
    \label{fig:Tt}}
\end{figure}

Note that the Courant-Friedrichs-Lewy (CFL) criterion for stability of
the leapfrog scheme \citep{Durran99},
\begin{eqnarray*}
  \mu^*\ \le\ \frac{1}{\pi}, 
\end{eqnarray*}
can sometimes be exceeded in the middle of a run, even though the
simulation is stable at $t = 0$ (cf., run E3a in
Figure~\ref{fig:cour}).  This is because the advective time-stepping
limit depends on the maximum speed $c_{\rm max}$, which can increase
during the evolution of a flow.  Careful monitoring of the physical
fields shows a zone of intense shear between the two vortices
generating small-scale oscillations that rapidly amplify until the
simulation becomes nonsense.  The culprit is not lack of spatial
resolution or a blocked turbulent cascade, in this case, because the
calculation can be extended indefinitely by halving the timestep.

Figure~\ref{fig:eps} illustrates the sensitivity of the evolution to
$\epsilon$ for the range, $\epsilon \in [0.01, 0.1]$.  Although the
resolution in these calculations is only T21, they illustrate the
point.  Snapshots of the flow field are shown at three successive
times for three simulations, differing only in the value of
$\epsilon$.  Similar flow patterns emerge in all the simulations: they
all exhibit a cyclic behavior with vortices translating around the
planet, undergoing large variations in strength and size as they do
so, with corresponding changes in the temperature field.
However, {\it at a given instant} the flow and temperature fields look
different between the three runs.  At $t = 130\,\tau_p$, in all the
runs there is a warm cyclone pair centered west of the substellar
point.  And in all the runs, the cyclones move westward and the flow and
temperature fields undergo substantial changes before eventually
returning to a similar state, 15--20 planet rotations later.  But at $t =
145\,\tau_p$, the run with the largest $\epsilon$ has already returned
to a state similar to that at $t = 130\,\tau_p$, while the runs with
smaller $\epsilon$ take longer to complete their cycles.

Figure~\ref{fig:Tt} shows the behavior more clearly.  The temperature
at a point on the model planet atmosphere (0$^{\circ}$ longitude,
30$^{\circ}$ latitude) evolves in time for two simulations which have
identical parameters, except for $\epsilon$.  The two runs match
nearly exactly until about 45$\,\tau_p$, when the two runs start to
deviate.  In the beginning only slightly.  But, at about 70$\,\tau_p$
the temperature oscillations in the run with the larger
$\epsilon$ lead in phase, compared to the run with smaller $\epsilon$.
This behavior agrees qualitatively with Equation~\ref{eq:phase}.  Over
long timescales the three simulations exhibit very similar behavior,
even if amplitudes, phases, and periodicities of the flow and
temperature fields are not exactly the same.  As noted, simulations
shown in Figures~\ref{fig:eps} and \ref{fig:Tt} are at T21 resolution,
but with higher resolution deviations appear even earlier.

In \citet{Showmanetal09}, the MITgcm \citep{Adcroft04} is used.  In that
study, the model employs the third-order Adams-Bashforth method, 
which has some attractive properties \citep{Durran91}.  
However, the scheme does require an initialization phase in
which $\Psi^1$ and $\Psi^2$ are computed from the initial condition
$\Psi^0$ by some other procedure, such as the fourth-order Runge-Kutta
or a first- or second-order scheme with several short timesteps. It 
should be emphasized that---as {\it it is a major point of this paper}---the
main concern is usually adequate spatial resolution, especially in
problems with inherent small-scale phenomena, not the
time-integration.  A second- or even first-order time-integration
scheme can be perfectly adequate for many purposes.

\section{Conclusion}\label{sec:concl}

A major aim of this paper has been to shed light on some crucial
aspects for numerical modeling of atmospheric circulation on hot
extrasolar planets.  Here we have shown that, a spectral model
offers advantages in accuracy and diagnostics, given that the
higher-order field and wavenumbers are what's actually evolved.
However, all numerical models, including spectral models, have
limitations in how well they can represent physical reality.
Moreover, the models can easily be applied outside the realm of
``safe parameters'' and produce results that are nonsensical.  The
challenge is to properly test and identify the limits.  When numerical
artifacts appear, it is important to know how to deal with them and to
know when a simulation should be discarded.

In this paper we have shown that, for hot extrasolar planets
simulations with stationary forcing, there is a strong sensitivity to
the strength of applied artificial viscosity.  In addition, there is a
relation between the thermal relaxation time $\tauth$ and the
viscosity $\nu$: small $\tauth$'s lead to a large amount of
unphysical, grid-scale oscillations in the simulation, which forces
the use of excessive amounts of artificial viscosity to quench the
oscillations.  Hence, using a fixed strength of artificial viscosity
in a simulation with a large range of $\tauth$ in the model domain
(e.g., from about an hour to tens of days)---as done in many
simulations in the literature---inevitably produces flow and
temperature fields, which are either dominated by unphysical noise or
over-damping.  One may then wish to apply a spatially varying $\nu$,
but clearly this is then motivated by a numerical basis rather than a
physical one.
 
The proper values to use for the relaxation time (or variables needed
for realistic radiative transfer) are not known.  Based on the
findings in this work, calculations with extremely short $\tauth$'s
warrant further scrutiny.  Current GCMs may not be standing up too
well to this stressful test.  If, however, the short $\tauth$ are
really {\it physically} relevant, then another form of heating/cooling
parameterization or setup is needed.  This is not a criticism of the
Newtonian relaxation scheme, which in fact has been (and continues to
be) very useful for understanding basic mechanisms.  

One solution could be to spatially vary the $\nu$, as already discussed; 
but, this would lead to further complexity.  Even if direct radiative transfer 
is incorporated, one must ensure that the forcing is not too violent
or strong (large amplitude and short timescale).  Indeed, a comparison 
of our $\nu$ values, scaled appropriately for the Earth, shows that we have 
had to use $\nu$ values higher than that normally used in Earth studies 
\citep{Collinsetal04}.  
As discussed in \citep{Cho08}, if the radiative processes appear as 
practically instantaneous from the perspective of the flow, then an 
adiabatic approach is more appropriate.  Certainly from a numerical 
accuracy standpoint, as motivated by the present work, adiabatic and 
``gently forced'' calculations are useful as baselines.  Else, gradually 
ramping up heating and/or initializing simulations close to a balanced 
state is necessary \citep{ThraCho10a}.

GCMs of extrasolar planet atmospheres have great value in helping to
guide and interpret observations.  It is then important to critically
examine the effects of the numerous parameters that are specified.
This is particularly crucial when applying the models to a ``new
regime'', where the physical conditions differ markedly from a
traditional (e.g., Earth) one.  In this paper we have shown examples
of how a commonly-used forcing can steer GCMs to produce misleading
results and how numerical expediencies, such as the Robert-Asselin
filter, can produce slewing frequency as well as the well-known
damping and phase-errors \citep{Durran91,Williams09}.  In addition, we
have discussed diagnostics procedures to better assess the quality
of a simulation using the vorticity field and energy spectra.
Reliance on spatial and temporal averages can effectively conceal
telltale signs that a simulation is not trustworthy.  

A simulation which is properly resolving the flow should approximately
conserve energy for a long time.  This should be so even if this
property is not explicitly built into the discretization algorithm as
in the scheme of \citet{Arakawa66} [this scheme conserves the
domain-integrated energy and enstrophy ($\frac{1}{2}\zeta^2$) in
the nonlinear advection term].  For only then can we trust that a
calculation is not artificially driven to an unphysical region in the
solution space.

\acknowledgments

The ideas and development of this work has had a long gestation
period.  The authors would like to thank O. M. Umurhan, C. Watkins,
A. G\"{u}l\c{s}en, I. Polichtchouk, and E. Staehling for helpful
discussions over the past several years.  The authors also
acknowledge the hospitality of the Kavli Institute for Theoretical
Physics, Santa Barbara, where some of the work was accomplished.
H.\TH.\TH.\ is supported by the EU Fellowship.  J.Y-K.C. is supported
by the STFC grant PP/E001858/1.


\end{document}